\documentclass[lettersize,journal]{IEEEtran}
\usepackage{amsmath,amsfonts}
\usepackage{algorithmic}
\usepackage{algorithm}
\usepackage{array}
\usepackage{textcomp}
\usepackage{stfloats}
\usepackage{url}
\usepackage{verbatim}
\usepackage{graphicx}
\usepackage{cite}
\usepackage{amsmath}
\usepackage{amsthm} 
\usepackage{multirow}
\usepackage{subcaption}
\usepackage{tcolorbox}
\usepackage{tikz}
\usetikzlibrary{arrows.meta, positioning, shapes.geometric, decorations.pathmorphing, calc}

\tikzstyle{phase} = [draw, rounded corners, minimum width=3.5cm, minimum height=4.5cm, thick]
\tikzstyle{process} = [rectangle, draw, minimum width=2.5cm, minimum height=0.8cm, fill=white]
\tikzstyle{arrow} = [thick, ->, >=Stealth]
\tikzstyle{circlepoint} = [circle, draw, fill=orange!70, minimum size=0.3cm, inner sep=0pt]

\usepackage{enumitem} 
\usepackage{xcolor}  
\usepackage{soul}    
\usepackage{stfloats}
\usepackage{booktabs}
\usepackage{framed}
\setulcolor{blue}
\theoremstyle{definition}
\usepackage{mathrsfs}

\usepackage{caption}
\usepackage{subcaption}
\usepackage{tikz}
\usetikzlibrary{positioning}
\usepackage{amssymb}

\usepackage[
    colorlinks=true,
    linkcolor=black,      
    citecolor=black,      
    urlcolor=blue         
]{hyperref}

\begin{document}

\title{ RIS Codebook Index Assignment under Imperfect Control Links Using TSP-Inspired Optimization}

\author{Liangshun Wu, Wen Chen, Qingqing Wu, Xudong Bai and Kunlun Wang 

\thanks{Liangshun Wu. Wen Chen, and Qingqing Wu are with the Department of Electronic Engineering, Shanghai Jiao Tong University, Shanghai
200240, China (e-mail: wuliangshun@sjtu.edu.cn; wenchen@sjtu.edu.cn; qingqingwu@sjtu.edu.cn).}
\thanks{Xudong Bai is with the College of Microelectronics, Northwestern Polytechnical University, Xi’an 710129, China (e-mail: baixudong@nwpu.edu.cn)}
\thanks{Kunlun Wang is with the School of Communication and Electronic Engineering, East China Normal University, Shanghai 200241, China (e-mail: klwang@cee.ecnu.edu.cn).}
}



\maketitle

\begin{abstract}
Reconfigurable Intelligent Surfaces (RIS) promise transformative gains in wireless communications by enabling programmable control of the propagation environment through discrete phase configurations. In practical deployments, the control of RIS phase states is typically managed using finite codebooks, with configuration indices transmitted over low-latency, yet imperfect, wireless feedback channels. Even rare feedback bit errors can lead to significant mismatches between intended and applied RIS states, degrading system performance. This paper addresses the challenge of robust RIS codebook index assignment by formulating it as a combinatorial optimization problem, equivalent to the Traveling Salesman Problem (TSP), where codewords are "cities" and edge weights reflect SNR degradation under codeword confusion. A novel three-phase heuristic algorithm is proposed to solve this, consisting of a provision phase, a shotgun phase, and a fuzzy concatenation phase. Simulation results show that the method outperforms conventional indexing strategies and achieves near-optimal robustness to index errors, while also being scalable and hardware-agnostic for real-time deployment. Future work includes multi-bit error correction and online adaptive mapping for time-varying channels.

\end{abstract}


\section{Introduction}

\IEEEPARstart{R}econfigurable Intelligent Surfaces (RIS) have emerged as a transformative technology for next-generation wireless networks, enabling programmable control of the radio environment to enhance communication performance. By dynamically adjusting the phase shifts of a large array of passive reflecting units (PRUs), an RIS can shape the propagation of electromagnetic waves to improve signal quality and suppress interference. As a result, RIS has been explored for boosting coverage, capacity, and energy efficiency in a wide range of scenarios, including multi-antenna and cell-free MIMO networks\cite{wang2025efficient,chen2025spatial,chen2024intelligent,gao2024irs}, interference-limited and wireless power transfer systems\cite{gao2024irs,gao2023exploiting}, and multi-user communication networks with advanced multiple access techniques\cite{zhang2024fairness,qiu2023intelligent}. Furthermore, RIS play a key role in emerging integrated sensing and communication applications\cite{zhao2024multi,zhang2024intelligent,meng2024intelligent}, facilitating functionalities such as joint radar-communication and interference cancellation\cite{peng2024semi,liu2024rate,hua2023secure}. In highly dynamic environments like vehicular networks, deploying RIS can significantly improve the reliability and throughput of wireless links\cite{qi2024reconfigurable,qi2024deep,meng2023sensing} by smartly reconfiguring the propagation channel in real time. These studies demonstrate the immense potential of RIS to enhance wireless communication systems when the RIS is properly configured based on the instantaneous channel conditions. However, realizing these gains in practice requires accurate channel state information (CSI) at the controlling base station (BS) and, critically, a reliable control link for updating the RIS—an aspect that has received relatively little attention to date.

\subsection{RIS Codebook Configuration and Feedback Channel Challenges}
RIS-assisted communication systems typically adopt a two-stage cascaded channel estimation: estimating BS-RIS and RIS-user channels separately, then cascading via RIS reflection matrix design into an equivalent channel\cite{zegrar2020general,wei2021channel}. To reduce pilot and computational overhead, methods like compressive sensing\cite{lin2022channel}, atomic norm minimization\cite{schroeder2022channel}, and deep learning\cite{kundu2021channel} are widely used. The BS selects an appropriate RIS configuration from a finite, pre-designed codebook based on current CSI and sends the corresponding index to the RIS controller via a dedicated feedback channel. Although many works implicitly assume this link is error-free and instantaneous, practical wireless control channels inherently introduce errors. Even minor bit errors in the index can significantly disrupt the intended RIS configuration, severely degrading system performance. While prior research addresses CSI overhead\cite{zhang2024often} and inaccuracies\cite{zhu2023reconfigurable}, the reliability of the RIS control link remains largely unexplored and critically important for practical deployments.

\subsection{Index Assignment (IA) Problem}
On top of the above scenario, the critical design problem is how to assign the codebook indices so as to maximize the robustness of RIS-assisted links to feedback errors. Specifically, given a pre-designed codebook shared between the BS and RIS controller, the assignment of binary indices to codewords fundamentally determines the system's sensitivity to index errors over the control channel. Since, in practical high-mobility or vehicular applications, the RIS must be reconfigured frequently and the feedback index is typically transmitted over a binary symmetric channel (BSC), the mapping strategy of indices to codewords directly impacts the end-to-end signal quality. The objective of the IA is to ensure that the most probable index errors—such as single-bit flips in the binary index—result in the least possible degradation of RIS-assisted performance. In mathematical terms, this is a combinatorial optimization problem: one aims to minimize the expected performance loss (e.g., SNR degradation) caused by index mismatch, given the error statistics of the feedback channel and the similarity/distance between codewords in the codebook. The IA problem can, in general, be formulated as a quadratic assignment problem (QAP), which accounts for all possible index confusions weighted by their respective error probabilities. QAPs are NP-complete, and even approximate solutions become intractable for large codebooks~\cite{zeger1990pseudo, farvardin1990study, knagenhjelm2002hadamard, spira2000codebook}. Several classical suboptimal algorithms have been proposed for QAP-based index mapping, such as the Binary Switching Algorithm (BSA)~\cite{zeger1990pseudo}, simulated annealing~\cite{farvardin1990study}, and the Linear Gain Switching Algorithm (LISA)~\cite{knagenhjelm2002hadamard}, as well as graph-based heuristics~\cite{spira2000codebook}. However, in practice, for small bit error rate $q$ (i.e., in the high BSC SNR regime), the probability of multiple simultaneous bit errors is negligible, so single-bit flips dominate the average loss calculation. In this case, the index assignment problem can be simplified:  adjacent indices in Hamming distance (i.e., single-bit different binary codes) become the dominant error pattern. Thus, the mapping of codebook indices can be modeled as finding an ordering such that neighboring indices (in the Hamming sense) correspond to codewords that are as similar as possible.

\subsection{Contributions}
This work addresses the IA problem by assigning RIS codebook indices such that single-bit feedback errors minimally degrade system performance. We propose reordering codewords along an optimized path, assigning indices using a Gray code sequence, thus ensuring minimal performance loss for single-bit errors. Specifically, we cast this codebook ordering task as a Traveling Salesman Problem (TSP): each codeword represents a ``city,'' and distances between cities reflect performance loss when codewords are mismatched. The shortest Hamiltonian path through this graph yields the optimal ordering minimizing SNR degradation.

Given the NP-hard nature of the TSP, exact solutions are impractical for larger codebooks. Hence, we propose a customized three-phase heuristic algorithm specifically tailored for RIS index assignment: (i) layered candidate pruning, (ii) stochastic ``shotgun'' route sampling, and (iii) fuzzy concatenation of high-frequency route segments. Our approach efficiently finds near-optimal index permutations, significantly outperforming random or natural assignments and achieving near-optimal performance with drastically lower computational cost and memory requirements than conventional TSP solvers.

The key contributions are summarized as follows:
\begin{itemize}
\item Formulating RIS codebook IA as a TSP, defining distances based on performance loss between RIS configurations. The shortest Hamiltonian path minimizes SNR degradation under single-bit errors.
\item Developing a fast, three-phase heuristic tailored for IA, outperforming standard methods, reducing computational complexity, and enabling practical RIS deployment.
\end{itemize}

The remainder of this paper is organized as follows. Section II introduces the system model and formulates the IA problem as a TSP. Section III details the proposed heuristic algorithm. Section IV presents performance analysis and simulation results. Section V concludes and discusses future research directions.

\begin{figure*}[htbp]
  \centering
  \includegraphics[width=\textwidth]{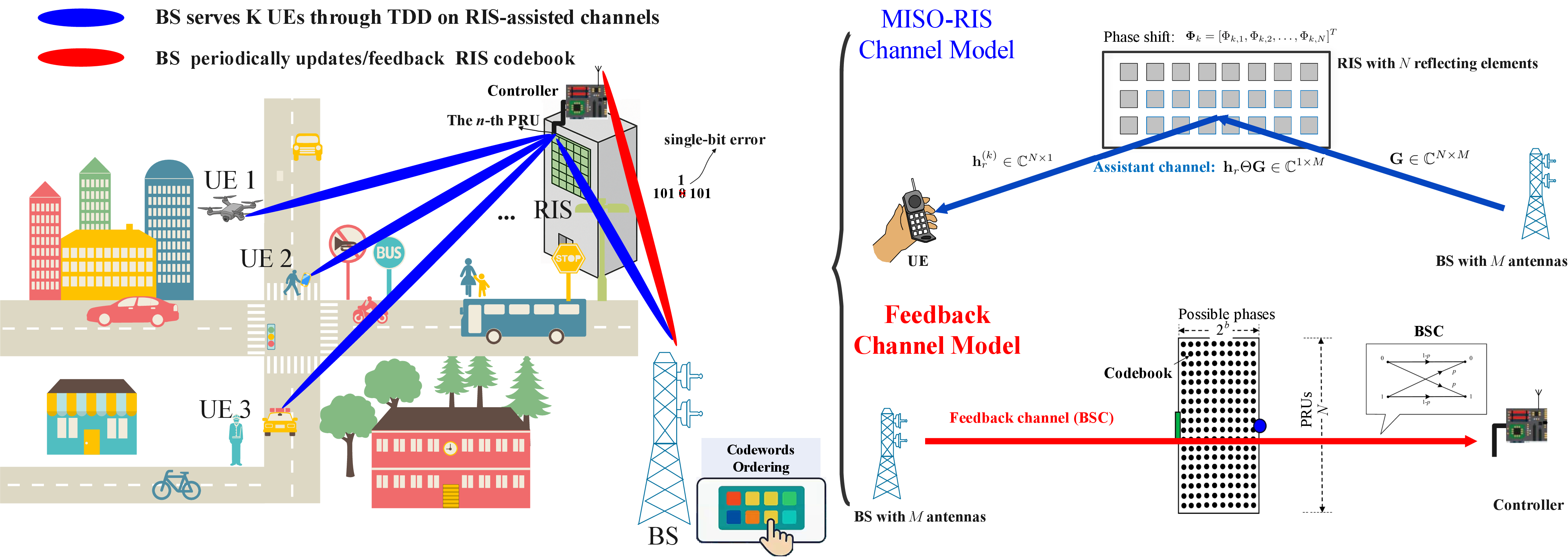}
  \caption{RIS-assisted downlink MISO communication scenario. }
  \label{scenario}
\end{figure*}
\begin{figure*}[h]
  \centering
  \includegraphics[width=\textwidth]{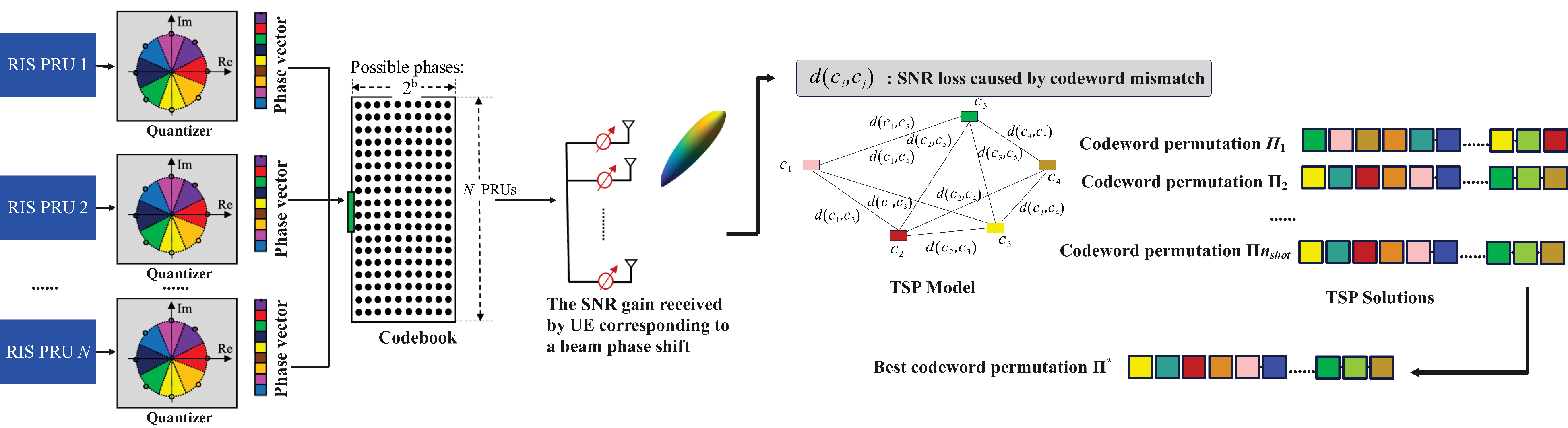}
  \caption{RIS codebook quantization and IA to TSP formulation. }
  \label{codebook}
\end{figure*}

\section{ System Model} \label{sec:model}

Consider an RIS-assisted downlink MISO communication scenario, as depicted in Fig.~\ref{scenario}. The BS, equipped with $M$ antennas, serves $K$ UEs through TDD (Time Division Duplexing). The BS broadcasts data to UEs both through a direct path and via a RIS of $N$ PRUs. This is a MISO scenario, where ``MISO'' stands for ``Multiple-Input Single-Output''. In this scenario, the BS, acting as the transmitter, is typically equipped with multiple antennas (enabling multiple inputs), while the UEs, as receivers, generally use a single antenna (enabling single output). 

A finite-size codebook of RIS phase shift configurations is shared between the BS and the RIS controller.  The CSI of RIS-assisted channels can be estimated using methods proposed in \cite{zegrar2020general,wei2021channel}. The BS dynamically selects the optimal codeword index from the codebook in real time. This index is then fed back to the RIS controller through a binary symmetric channel (BSC), where each bit may be flipped with a certain probability—typically resulting in single-bit errors. The controller updates the RIS configuration according to the received (possibly erroneous) index. As a result, the RIS may activate a phase configuration different from the intended one, leading to index mismatch distortion and degraded communication performance.

The RIS configuration must be updated periodically to track channel changes, particularly in dynamic environments like vehicular networks. If we assume that the RIS controller updates the phase shift configuration every 1 second, the maximum positional offset for a vehicle moving at a speed of 60 km/h (16 m/s) would be approximately 16 meters in that time. This small deviation due to periodic codebook updates is typically acceptable, as the RIS configuration is designed to handle such slight changes in UE position without significantly affecting the overall system performance.

\begin{table}[h]
\centering
\caption{List of Abbreviations}
\begin{tabular}{cl}
\hline
{Abbreviation} & {Full Form} \\
\hline
RIS & Reconfigurable Intelligent Surface \\
BS & Base Station \\
CSI & Channel State Information \\
UE & User Equipment \\
SNR & Signal-to-Noise Ratio \\
MISO & Multiple-Input Single-Output \\
PRU & Passive Reflecting Unit \\
TDD & Time Division Duplexing\\
BSC & Binary Symmetric Channel \\
TSP & Traveling Salesman Problem \\
IA & Index Assignment \\
QAP & Quadratic Assignment Problem \\
\hline
\end{tabular}
\label{tab:abbreviations}
\end{table}

The abbreviations used in the system model are summarized in Table~\ref{tab:abbreviations}.

\subsection{MISO–RIS Model}

For any given UE (say UE $k$),  the direct channel from the BS to the UE is blocked;  let $\mathbf{h}_r^{(k)} \in \mathbb{C}^{N\times1}$ the channel from the RIS to the UE, and $\mathbf{G} \in \mathbb{C}^{N\times M}$ the channel from the BS to the RIS. Under narrowband flat-fading, the combined downlink channel for UE $k$ is:
\begin{equation}
\mathbf{h}^{(k)} = \mathbf{G}^T \Theta_k\, \mathbf{h}_r^{(k)},
\end{equation}
where $\Theta_k = \mathrm{diag}(\boldsymbol{\Phi}_k)$ is the RIS reflection matrix for UE $k$, and 
\begin{equation}
\boldsymbol{\Phi}_k = [\Phi_{k,1}, \Phi_{k,2}, \dots, \Phi_{k,N}]^T
\end{equation}
is the vector of phase shifts applied by the $N$ RIS PRUs (with each $\Phi_{k,n} \in E$, $E$ is defined in eq. (\ref{eq:e})).

\subsection{IA as a TSP Problem}
Fig. \ref{codebook} (Left) provides an overview of the codebook-based RIS system and the index assignment optimization. Left (Codebook Generation): Each RIS PRU applies $b$-bit quantized phase shifts over $[0, 2\pi)$ --- i.e., the set of allowable discrete phase values is
\begin{equation}
E = \{ e^{j0},\, e^{j\Delta\theta},\, \dots,\, e^{j(2^b-1)\Delta\theta} \},
\label{eq:e}
\end{equation}
where $\Delta\theta = 2\pi / 2^b$.

A finite codebook $\mathcal{C} = \{\mathbf{c}_0, \mathbf{c}_1, \dots, \mathbf{c}_{K-1}\}$ of $K$ candidate RIS configurations is designed (e.g., through beamsteering or quantization codebook methods). Each codeword $\mathbf{c}_i$ is an $N$-dimensional vector of quantized phase shifts, 
\begin{equation}
\mathbf{c}_k = \boldsymbol{\Phi}_k= [\Phi_{k,1}, \dots, \Phi_{k,N}]^T,
\end{equation}
and can be interpreted as a particular phase configuration across the RIS. In practice, this codebook might be shared by all users or tailored per user; in our scenario, each time slot the BS selects one codeword from $\mathcal{C}$ to serve the currently scheduled user. We assume time-division user scheduling, so the RIS serves one user at a time and updates its configuration each slot accordingly.

For RIS configuration $\mathbf{c}_i$ (i.e., $\Theta_i = \mathrm{diag}(\mathbf{c}_i)$), the received SNR at UE $k$ is defined as
\begin{equation}
\mathrm{SNR}_i = \frac{P, |\mathbf{h}^{(k)}_i|^2}{\sigma^2},
\end{equation}
where $\mathbf{h}^{(k)}_i = \mathbf{G}^T \Theta_i \mathbf{h}_r^{(k)}$ is the effective downlink channel for UE $k$ under configuration $\mathbf{c}_i$, $P$ is the transmit power, and $\sigma^2$ is the noise variance.

The BS uses the available CSI to choose the index $i$ of the optimal RIS codeword for the active UE (to maximize the received SNR, for instance). The BS then feeds back this index to the RIS controller over a limited-rate feedback link. We model the feedback channel as a binary symmetric channel with bit error probability $q \ll 1$. Thus, each feedback index is represented by a $\log_2 K$-bit sequence, and each bit may be flipped independently with probability $q$. We focus on the dominant error events of a single-bit flip and neglect multi-bit errors (since $q$ is very small). In other words, if the intended index is $i$, it might be received as $j$ with probability $q$ if $i$ and $j$ differ in exactly one bit (Hamming distance 1), and with negligible probability if the Hamming distance is greater than 1. We also assume all codewords are equally likely to be chosen (uniform prior on $i$).

Each codeword $\mathbf{c}_k$ may correspond to a different effective SNR when applied (depending on the UE's channels $\mathbf{h}_d^{(k)}, \mathbf{h}_r^{(k)}$). Thus, if a feedback error causes the index $i$ to be mistaken as $j$, RIS configuration will be $\mathbf{c}_j$ instead of $\mathbf{c}_i$. This results in a performance drop due to the mismatch. We define the mismatch loss between codewords $\mathbf{c}_i$ and $\mathbf{c}_j$ as the relative SNR loss:
\begin{equation}
d(\mathbf{c}_i,\mathbf{c}_j) = \Big|1 - \frac{\mathrm{SNR}_j}{\mathrm{SNR}_i}\Big|,
\end{equation}
where $\mathrm{SNR}_i$ is the received SNR when the RIS uses configuration $\mathbf{c}_i$. For instance, if $\mathrm{SNR}_j$ is much lower than $\mathrm{SNR}_i$, then $d(\mathbf{c}_i,\mathbf{c}_j) \approx 1$ (significant loss), whereas if $\mathrm{SNR}_j$ is close to $\mathrm{SNR}_i$, then $d(\mathbf{c}_i,\mathbf{c}_j)$ is small. (In practice, one could also measure the loss in absolute terms $|\mathrm{SNR}_i - \mathrm{SNR}_j|$ or in decibels $10\log_{10}(\mathrm{SNR}_i/\mathrm{SNR}_j)$; we use the relative definition for convenience.)

Our goal is to assign binary indices to the RIS codewords (or equivalently, to determine an ordering of the codewords in the codebook) such that the average SNR loss due to a single-bit index error is minimized. Formally, the objective is to minimize
\begin{equation}
\mathbb{E}[d] \;=\; \sum_{i=0}^{K-1}\sum_{j=0}^{K-1} P_i\,P_{j|i}\; d(\mathbf{c}_i,\mathbf{c}_j),
\end{equation}
where $P_i = 1/K$ and $P_{j|i}$ is the probability that index $i$ is decoded as $j$ at the RIS. Under our single-bit error assumption, $P_{j|i} \approx q$ if $\mathrm{Ham}(i,j)=1$ (indices differ by one bit) and $P_{j|i}\approx 0$ otherwise. Plugging these in, and noting $q$ and $K$ are constants, the optimization simplifies to minimizing
\begin{equation}
\sum_{i=0}^{K-1}~\sum_{\substack{j:~\mathrm{Ham}(i,j)=1}} d(\mathbf{c}_i,\mathbf{c}_j),
\end{equation}
the total perturbation between all pairs of codewords whose indices differ in one bit.

\begin{table}[h]
\centering
\caption{Example of RIS codebook index assignment for $K=16$ codewords, where each codeword consists of $N=2$ phases, each quantized to $\{0^\circ, 90^\circ, 180^\circ, 270^\circ\}$ ($b=2$ bits).}
\begin{tabular}{ccc}
\hline
Gray Code & Codebook Index & Codeword\\
\hline
0000 & 6  & $[90^\circ,\,180^\circ]$ \\
0001 & 2  & $[0^\circ,\,180^\circ]$ \\
0011 & 7  & $[90^\circ,\,270^\circ]$ \\
0010 & 3  & $[0^\circ,\,270^\circ]$ \\
0110 & 14 & $[270^\circ,\,180^\circ]$ \\
0111 & 10 & $[180^\circ,\,180^\circ]$ \\
0101 & 15 & $[270^\circ,\,270^\circ]$ \\
0100 & 11 & $[180^\circ,\,270^\circ]$ \\
1100 & 4  & $[90^\circ,\,0^\circ]$ \\
1101 & 0  & $[0^\circ,\,0^\circ]$ \\
1111 & 5  & $[90^\circ,\,90^\circ]$ \\
1110 & 1  & $[0^\circ,\,90^\circ]$ \\
1010 & 12 & $[270^\circ,\,0^\circ]$ \\
1011 & 8  & $[180^\circ,\,0^\circ]$ \\
1001 & 13 & $[270^\circ,\,90^\circ]$ \\
1000 & 9  & $[180^\circ,\,90^\circ]$ \\
\hline
\end{tabular}
\label{example}
\end{table}

This problem can be mapped to a Traveling Salesman Problem (see Fig. \ref{codebook} (Right)). We construct a complete graph where each node corresponds to a codeword, and the weight of the edge between node $i$ and node $j$ is given by $d(\mathbf{c}_i,\mathbf{c}_j)$. Our goal is to find a path that visits each node (codeword) exactly once and has the minimum possible sum of edge weights. If we find a Hamiltonian path through the graph that minimizes the total weight, then assigning the codewords in that order (and labeling them sequentially or via Gray code) will ensure that any single-bit index error leads to jumping to an adjacent codeword on this path — which by optimality of the path is a configuration with minimal possible loss relative to the intended one. In effect, the ``distance'' between any two codewords that are one bit apart in the new index assignment is as small as possible.

Mathematically, let $\pi = [\pi(0), \pi(1), \dots, \pi(K-1)]$ be a permutation of $\{0,1,\ldots,K-1\}$ representing an ordering of the $K$ codewords. The total loss along this ordering (viewed as a path) is
\begin{equation}
L(\pi) = \sum_{k=0}^{K-2} d\!\Big(\mathbf{c}_{\,\pi(k)},~ \mathbf{c}_{\,\pi(k+1)}\Big).
\end{equation}
(For an open Hamiltonian path, we do not include the edge from the last back to the first.) We seek the permutation $\pi^*$ that minimizes $L(\pi)$. The optimal index assignment will then arrange the codewords in the order $\pi^*$ and use a labeling (such as a binary Gray code) that ensures adjacent codewords on the path differ by one bit in their indices.

As a simple example, consider a case with $N=2$ RIS PRUs and $b=2$ quantization bits per PRU, yielding $K = 16$ codewords. Each codeword is a 2-dimensional phase vector with each PRU in $\{0^\circ,90^\circ,180^\circ,270^\circ\}$. We apply the following procedures:
\begin{enumerate}[label=Step\arabic*:]
\item Solve the TSP on the 16 codewords to obtain the optimal permutation $\pi^* = [\pi(0), \pi(1), \ldots, \pi(15)]$. Here, $\pi(x)$ denotes the original codebook index (0-15) of the codeword at position $x$ along the optimal path.
\item Assign Gray code labels sequentially along the TSP path: label position $k$ with the $k$-th 4-bit Gray code $g_k$ (where $g_0=0000$, $g_1=0001$, $\dots$, $g_{15}=1111$), and relabel each codeword in the codebook according to its Gray code assignment. Thus, the codeword at position $k$ in the optimal path $\pi^*$ is given the index $g_k$ in the new codebook mapping.
\end{enumerate}

Table \ref{example} illustrates this example. It lists the Gray code assigned to each codeword, the original index of that codeword (from the unsorted codebook), and the corresponding phase vector. By construction, consecutive Gray codes differ in only one bit, and because adjacent codewords on the path $\pi^*$ were chosen to have small SNR loss, a single-bit error in the index will likely select a codeword with a similar phase configuration and thus a small performance loss.

\section{The Proposed Algorithm }\label{sec:method}

\begin{figure*}
\includegraphics[width=\linewidth]{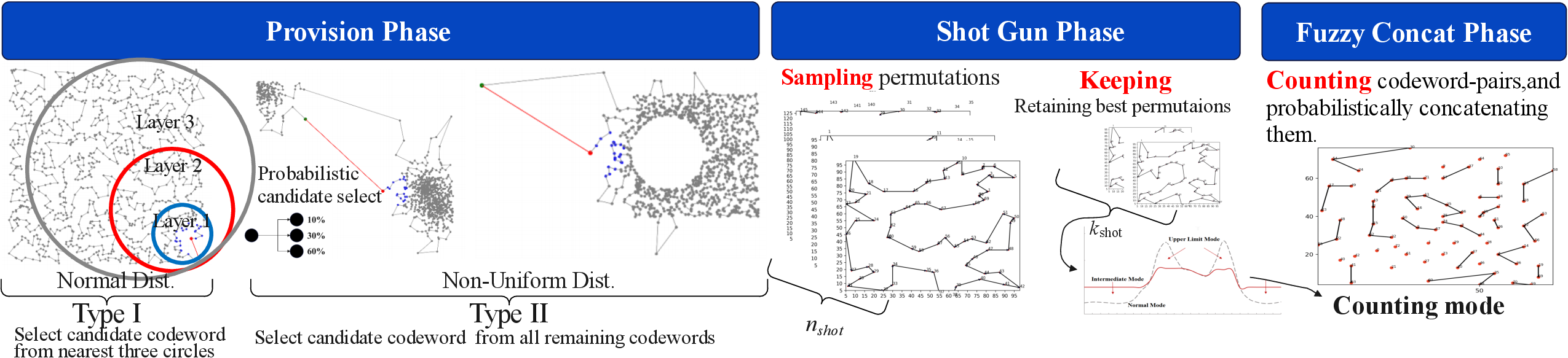}
\caption{The proposed TSP solver.}
\label{pro}
\end{figure*}

To efficiently find a near-optimal index permutation for the RIS codebook, we propose a novel TSP solver. Fig. \ref{pro} provides a conceptual overview of the algorithm, which consists of: (1) a \textit{Provision Phase} that identifies promising candidate moves based on the distribution of codeword perturbations, (2) a \textit{Shotgun Phase} that generates and refines a large number of candidate permutations using biased random sampling, and (3) a \textit{Fuzzy Concatenation Phase} that iteratively fine-tunes the best permutations by reinforcing frequently observed good partial routes. Pseudocode for the entire algorithm is given in Algorithm \ref{alg:tsp_three_phase}, and default parameter values are listed in Table \ref{tab:defaults}.   The proposed solver’s polynomial time complexity is formally proved in Appendix A.

We next describe each phase in detail.

\subsection{ Provision Phase}

In this initial phase, we analyze the pairwise perturbation values $d(\mathbf{c}_i,\mathbf{c}_j)$ for the given codebook to guide the subsequent search. We classify the distribution of these pairwise losses into two types:
\begin{itemize}
\item Type I: The distribution of $d(\mathbf{c}_i,\mathbf{c}_j)$ (for all pairs) is approximately unimodal or single-peaked (often roughly normal).
\item Type II: The distribution is multi-modal or otherwise irregular (non-normal).
\end{itemize}

If the problem is Type I, it means most codeword pairs have moderate loss and truly large losses are rare outliers; this suggests a strong notion of ``neighborhood'' among codewords. In this case, we will restrict our search to favor moves to the closest remaining codewords first. For Type II, such a neighborhood structure is weaker, so a broader search is allowed at each step. 

To implement this, for each codeword $\mathbf{c}_i$ we precompute an ordered list of all other codewords sorted by the loss $d(\mathbf{c}_i,\cdot)$ (from smallest loss to largest). We then define circle-layer neighbor sets for $\mathbf{c}_i$ as follows: the first layer $L_i^{(1)}$ contains the $l_1$ closest codewords to $\mathbf{c}_i$, the second layer $L_i^{(2)}$ contains the next $l_2$ closest codewords, and the third layer $L_i^{(3)}$ contains the next $l_3$ closest. Here $l_1$, $l_2$, $l_3$ are preset numbers (with $l_1 < l_2 < l_3$) that depend on $K$ (see Table \ref{tab:defaults} for typical values). Intuitively, $L_i^{(1)}$, $L_i^{(2)}$, $L_i^{(3)}$ partition the set of nearest neighbors of $\mathbf{c}_i$ into three concentric ``circles'' of increasing radius (perturbation).

As we construct a path through the codebook, let $\mathcal{R}$ denote the set of codewords already included (the route so far), and let $\mathcal{C}$ be the set of all $K$ codewords. The remaining set of codewords at a given step (starting from current codeword $\mathbf{c}_i$) is simply $\mathcal{C} \setminus \mathcal{R}$, the set of codewords not yet visited.

Now we define the candidate set $\Omega_i$ for the next move from $\mathbf{c}_i$. If the problem is Type I (unimodal distribution), we choose $\Omega_i$ in a hierarchical, greedy manner:
\begin{itemize}
\item Layer-1 preference: If any codeword in $L_i^{(1)}$ has not been visited yet (i.e., $L_i^{(1)} \cap (\mathcal{C}\setminus\mathcal{R}) \neq \emptyset$), then let $\Omega_i = L_i^{(1)} \cap (\mathcal{C}\setminus\mathcal{R})$. (Choose from the closest neighbor layer.)
\item Layer-2 fallback: Otherwise, if no first-layer neighbors remain unvisited, then let $\Omega_i = L_i^{(2)} \cap (\mathcal{C}\setminus\mathcal{R})$ (if this is non-empty).
\item Layer-3 fallback: If neither first- nor second-layer neighbors are available, then let $\Omega_i = L_i^{(3)} \cap (\mathcal{C}\setminus\mathcal{R})$, provided this set is non-empty.
\item Last resort: If all of $L_i^{(1)}$, $L_i^{(2)}$, $L_i^{(3)}$ have been exhausted (or if $l_3$ was set such that these three layers don't cover all codewords), then we set $\Omega_i$ to the entire remaining set $\mathcal{C}\setminus\mathcal{R}$.
\end{itemize}

On the other hand, for Type II problems (no clear single peak in the distribution), we do not enforce a layered restriction. In that case, we simply take $\Omega_i = \mathcal{C}\setminus\mathcal{R}$, i.e., any of the remaining codewords can be the next candidate with equal structural priority.

This \textit{Provision Phase} ensures that for ``well-behaved'' instances (Type I), the algorithm will prioritize transitions to near-by codewords (in terms of SNR loss) before considering farther ones, which is a greedy bias that can reduce the total path cost. For more complex instances (Type II), the algorithm remains flexible to explore wider moves.

See the left panel of Fig.~\ref{pro} for a visual illustration of the \textit{Provision Phase} and its Type I/Type II classification.

\subsection{ Shotgun Phase}

In the \textit{Shotgun Phase}, we generate a large number of candidate Hamiltonian paths through the codewords using a randomized constructive approach, and then we filter and retain the best ones. The term ``shotgun'' reflects that we shoot out many random trials in hope of hitting near-optimal solutions, while using the \textit{Provision Phase} rules to guide each trial towards reasonable territory.

Sampling: We construct $n_{\text{shot}}$ random permutations of the $K$ codewords (with $n_{\text{shot}}$ on the order of $K^2$ according to Table \ref{tab:defaults}). Each permutation is built step-by-step as follows:
\begin{enumerate}[label=Step\arabic*:]
\item Start: Randomly pick an initial codeword to start the path.
\item Iterative extension: Suppose the current codeword is $\mathbf{c}_i$. Determine the candidate set $\Omega_i$ of next codewords using the rules from the \textit{Provision Phase} (based on Type I or II and the remaining unvisited codewords). Then, for each candidate $\mathbf{c}_m \in \Omega_i$, compute a selection probability $p_{\mathbf{c}_m}$ according to a biased rule:
\begin{itemize}
   \item Compute 
\begin{equation}
\bar{d} = \frac{1}{|\Omega_i|}\sum_{\mathbf{c}_m \in \Omega_i} d(\mathbf{c}_i, \mathbf{c}_m),
\end{equation}
 the average loss to the candidates.
   \item Choose a perturbation adjustment coefficient $\mu$ (a constant between 0 and 1, initial value given in Table \ref{tab:defaults}).
   \item Assign
\begin{equation}     
p_{\mathbf{c}_m} = \frac{\xi}{\,1 + \Big(\frac{d(\mathbf{c}_i,\mathbf{c}_m)}{\mu\,\bar{d}}\Big)^2}\,,
\end{equation}
     where $\xi$ is a normalizing constant ensuring $\sum_{\mathbf{c}_m \in \Omega_i} p_{\mathbf{c}_m} = 1$.
     This probability formula is designed to favor moves that have a small loss $d(\mathbf{c}_i,\mathbf{c}_m)$ (relative to the average), while still allowing a non-zero chance for larger-loss moves. The parameter $\mu$ controls the greediness: a smaller $\mu$ makes the probability more sharply favor the best (smallest loss) candidate, whereas a larger $\mu$ flattens the probabilities, introducing more randomness.
\end{itemize}
\item Random choice: Randomly select the next codeword from $\Omega_i$ according to the probabilities ${p_{\mathbf{c}_m}}$. Move to that codeword and mark it visited. Repeat the process to pick subsequent codewords until all $K$ codewords have been visited.
\item Route finetuning: If at any point the number of unvisited codewords becomes very small (we set a threshold $f$, typically $f=4$), we switch to an exhaustive search to finalize the ordering of those remaining codewords optimally. In other words, when only $f$ or fewer codewords remain, we evaluate all $f!$ possible orderings of them (which is feasible for such small $f$) and choose the ordering that yields the minimum additional path cost. This ensures that the tail end of the path is locally optimal.
\item Closure: We obtain a complete route visiting all codewords. (If treating the problem strictly as a path and not a cycle, we do not need to return to the start; if a cycle were needed, we could close it now, but for index assignment an open path suffices.)

We perform the above procedure $n_{\text{shot}}$ times to produce $n_{\text{shot}}$ candidate permutations (each starting from a potentially different random codeword and following a different random trajectory guided by the probabilities).

\item Keeping: Among all the generated permutations, we retain the $k_{\text{shot}}$ shortest ones (those with the smallest $L(\pi)$). The values $n_{\text{shot}}$ and $k_{\text{shot}}$ are chosen such that $k_{\text{shot}} \ll n_{\text{shot}}$, providing a significant selection of the best candidates.

\item Counting: Next, we analyze the best $k_{\text{shot}}$ routes to gather statistics for the \textit{Fuzzy Concatenation Phase}. In particular, we count the frequency of each consecutive codeword pair among these top routes. If a pair of codewords $(\mathbf{c}_u, \mathbf{c}_v)$ appears consecutively in a route (regardless of order, i.e., either $u$ followed by $v$ or $v$ followed by $u$ along the path), we increment the count for that unordered pair. We ensure each pair is counted at most once per route to avoid biasing from a single route (though in a Hamiltonian path each unordered pair appears at most once anyway). These pair frequency counts $\delta_{(u,v)}$ will be used to influence the next phase.
\end{enumerate}
The process of permutation sampling and selection in the \textit{Shotgun Phase} is depicted in the center of Fig.~\ref{pro}.

\subsection{ Fuzzy Concatenation Phase}

The final phase aims to refine the solution by leveraging the historical frequency information of codeword pairs. The term "fuzzy concatenation" refers to the strategy of piecing together good partial routes (segments of consecutive codewords) in a probabilistic manner, allowing some flexibility (``fuzziness'') to escape local optima.

We iterate over a certain number of rounds (up to $T$ iterations, or until no improvement is observed). In each iteration, we generate a new set of $n_{\text{cate}}$ candidate routes (where ``cate'' stands for concatenation). The generation process is similar to the \textit{Shotgun Phase} but with two key differences:
\begin{enumerate}
\item We gradually reduce the perturbation coefficient $\mu$ as the iterations progress: $\mu \leftarrow \max\{\mu_0 - \sigma t, \mu_{\min}\}$ at iteration $t$, where $\mu_0$ is the initial value (same as used in the \textit{Shotgun Phase}), $\sigma$ is a small decay factor, and $\mu_{\min}$ is a lower bound. This means as we advance through iterations, the selection probabilities $p_{\mathbf{c}_m}$ become increasingly greedy (sharply favoring low-loss edges), thus focusing more on promising routes.
\item We adjust the selection probability using the pair frequency counts. Specifically, when determining the next hop from $\mathbf{c}_i$, for each candidate $\mathbf{c}_m \in \Omega_i$ we compute a historical weight:
   \begin{equation}
w_{\mathbf{c}_m} = \frac{\delta_{(i,m)}}{\sum_{\mathbf{c}_j \in \Omega_i} \delta_{(i,j)} + \epsilon}\,.
\end{equation}
   Here $\delta_{(i,m)}$ is the count of pair $(\mathbf{c}_i,\mathbf{c}_m)$ from the previous phase (or from previous iterations of the \textit{Fuzzy Concatenation Phase}), and $\epsilon$ is a tiny constant to avoid division by zero (if none of the candidates were seen before, we can set all $w_{\mathbf{c}_m} = 1/|\Omega_i|$ to have no bias). This weight $w_{\mathbf{c}_m}$ reflects how often $c_i$ and $c_m$ appeared consecutively in good paths. We then modify the selection probability by incorporating this weight:
\begin{equation}   
p_{\mathbf{c}_m} \;\leftarrow\; \frac{\,w_{\mathbf{c}_m} \cdot p_{\mathbf{c}_m}\,}{\sum_{\mathbf{c}_j \in \Omega_i} w_{\mathbf{c}_j} \cdot p_{\mathbf{c}_j}}\,.
\end{equation}
   In effect, if a candidate edge $(i,m)$ has been frequently used in high-quality solutions, it will get a higher probability, whereas rarely used or never-seen transitions get down-weighted (unless they have an intrinsically very low loss $d(i,m)$ that makes $p_{\mathbf{c}_m}$ large to begin with).
\end{enumerate}

Using the adjusted probabilities, we stochastically construct each of the $n_{\text{cate}}$ routes just as in the \textit{Shotgun Phase} (including the same fine-tuning step for the last few codewords). This yields a new pool of candidate solutions.

After generating these $n_{\text{cate}}$ routes in the current iteration, we again keep the best $k_{\text{cate}}$ among them (where $k_{\text{cate}}$ may be reduced gradually with each iteration to concentrate on the top routes, but is bounded below by $k_{\min}$ to maintain some diversity). We update the pair frequency counts $\delta_{(u,v)}$ based on these retained routes (adding to the counts from previous iterations).

To prevent the algorithm from getting stuck in a narrow subset of routes, we introduce a dynamic fuzziness control on the pair counts:
\begin{itemize}
\item Upper Limit Mode: If we detect that a small number of particular edges are appearing too frequently (dominating the routes), we cap their counts. Specifically, if a pair count $\delta_{(u,v)}$ exceeds a certain threshold (which can be a fraction of the total count $Q$ of all pairs seen so far in the iteration), we reset $\delta_{(u,v)}$ to that threshold. This prevents a few edges from overwhelming the probability bias.
\item Intermediate Mode: If many pairs have very similar counts in an intermediate range (for example, 40-80\% of the maximum count), it can indicate the algorithm is oscillating among similar routes. In this mode, we slightly reduce the counts in that range (e.g., scale them down to 20-40\% of $Q$) to encourage exploration of different combinations.
\item Normal Mode: By default, we use the raw counts as weights without additional filtering.
\end{itemize}

We monitor the progress over iterations. If the length of the best route found does not improve for a certain number of iterations, we switch the mode (e.g., from Normal to Upper Limit, or to Intermediate) to shake up the search. This adaptive adjustment of the weight counts adds a ``fuzzy'' element — we are willing to sometimes forget or downplay certain historical biases to potentially discover a better path. If we cycle through all modes and still observe no improvement in the best solution, we conclude that the algorithm has likely converged and terminate the search.

See the rightmost portion of Fig.~\ref{pro} for the visualization of the codeword-pair counting and adaptive mode switching in the \textit{Fuzzy Concatenation Phase}.

At the end of the \textit{Fuzzy Concatenation Phase}, the shortest path (permutation) found across all iterations is output as the final index assignment. This permutation can then be used to reindex the codebook (using Gray code labeling as described earlier). 

\begin{table}[htbp]
\caption{Default algorithm parameters for the Three-Phase TSP Solver.}\label{tab:defaults}
\centering
\begin{tabular}{lll}
\hline
Parameter & Default Value & Description \\
\hline
$l_1$                & $\approx \sqrt{K},\pm 1$  & Size of 1st neighbor layer (Type I)             \\
 $l_2$                & $\approx 2\sqrt{K},\pm 2$ & Size of 2nd neighbor layer (Type I)             \\
 $l_3$                & $K/3$                     & Size of 3rd neighbor layer (Type I)             \\
 $f$                   & 4                           & Exhaustive search threshold (remaining nodes)   \\
 $n_{\rm shot}$       & $3K^2$                    & Number of routes sampled in Shotgun Phase       \\
 $k_{\rm shot}$       & $3K$                      & Number of top routes kept from Shotgun Phase    \\
 $n_{\rm cate}$       & $200K$                    & Routes sampled per iteration in Fuzzy Phase    \\
 $k_{\rm cate}^{(0)}$ & $4K$                      & Initial number of top routes kept per iteration \\
 $\mu_0$              & 0.5                         & Initial perturbation coefficient                \\
 $\sigma$              & 0.01                        & Perturbation coefficient decay per iteration    \\
 $\mu_{\min}$         & 0.15                        & Minimum perturbation coefficient                \\
 $z$                   & $K/20$                    & Decay rate of $k_{\rm cate}$ per iteration   \\
 $k_{\min}$           & 200                         & Minimum routes to keep per iteration            \\
 $T$                   & $\sqrt{K}$                & Maximum number of Fuzzy Phase iterations        \\
\hline
\end{tabular}
\end{table}

\begin{algorithm}
\caption{Three-Phase TSP-Based Index Assignment}
\label{alg:tsp_three_phase}
\begin{algorithmic}[1]
\REQUIRE Perturbation matrix $D$; parameters: $l_1$, $l_2$, $l_3$, $f$, $n_{\mathrm{shot}}$, $k_{\mathrm{shot}}$, $n_{\mathrm{cate}}$, $k_{\mathrm{cate}}$, $\mu_0$, $\sigma$, $\mu_{\min}$, $z$, $k_{\min}$
\ENSURE Optimal codeword permutation $\pi^*$ minimizing total switching loss

\STATE \textbf{/*  Provision Phase: prepare neighbor layers for each codeword  */}

\FOR{each codeword $\mathbf{c}_i$}
    \STATE Sort all other codewords by $d(\mathbf{c}_i, \cdot)$ (relative SNR switching loss)
    \STATE Assign $l_1$ closest to $L_i^{(1)}$, next $l_2$ to $L_i^{(2)}$, next $l_3$ to $L_i^{(3)}$
\ENDFOR

\STATE \textbf{/*  Shotgun Phase: stochastic construction of many routes  */}  
\FOR{$n = 1$ to $n_{\mathrm{shot}}$}
    \STATE Randomly select start codeword; initialize route $\pi^{(n)}$
    \WHILE{not all codewords are visited}
        \STATE Determine candidate set $\Omega$ for next move using Provision Phase rules
        \FOR{each candidate $\mathbf{c}_j \in \Omega$}
            \STATE Compute $\bar{d}$, the mean loss to candidates
            \STATE Compute $p_{\mathbf{c}_j} = \frac{\xi}{1 + (d(\mathbf{c}_i, \mathbf{c}_j)/(\mu \bar{d}))^2}$ \COMMENT{Probability, $\xi$ normalizes $\sum p_{\mathbf{c}_j}=1$}
        \ENDFOR
        \STATE Sample next codeword from $\Omega$ according to $p_{\mathbf{c}_j}$
        \IF{number of unvisited codewords $= f$}
            \STATE Enumerate all $f!$ orders for final positions; choose minimum path cost
        \ENDIF
    \ENDWHILE
\ENDFOR
\STATE Retain $k_{\mathrm{shot}}$ shortest permutations (by total path cost)
\STATE Count consecutive codeword pairs (unordered) in retained permutations for frequency statistics

\STATE \textbf{/* Fuzzy Concatenation Phase: biased sampling based on pair statistics */}
\STATE Initialize edge (pair) frequency count matrix $\delta$
\FOR{iteration $t = 1$ to $T$ or until no improvement}
    \STATE Update $\mu \leftarrow \max\{\mu_0 - \sigma t,\, \mu_{\min}\}$
    \STATE $k_{\mathrm{cate}} \gets \max\{k_{\mathrm{cate}} - z t,\, k_{\min}\}$
    \FOR{$n = 1$ to $n_{\mathrm{cate}}$}
        \STATE Construct route using stochastic sampling as in Shotgun Phase, but:
        \FOR{each candidate $\mathbf{c}_j \in \Omega$}
            \STATE Compute pairwise frequency $w_{\mathbf{c}_j} = \frac{\delta_{(i,j)}}{\sum_{\mathbf{c}_j \in \Omega} \delta_{(i,j)} + \epsilon}$
            \STATE Adjust selection: $p_{\mathbf{c}_j} \gets \frac{w_{\mathbf{c}_j} \cdot p_{\mathbf{c}_j}}{\sum w_{\mathbf{c}_j} \cdot p_{\mathbf{c}_j}}$
        \ENDFOR
        \STATE Exhaustively search for optimal tail if unvisited $= f$
    \ENDFOR
    \STATE Retain $k_{\mathrm{cate}}$ best routes
    \STATE Update pair frequency counts $\delta_{(u,v)}$
    \STATE Dynamically switch frequency count mode (Normal/Intermediate/Upper Limit) if stagnation detected
    \IF{no improvement over predefined rounds}
        \STATE \textbf{break}
    \ENDIF
\ENDFOR
\RETURN Best permutation $\pi^*$ found
\end{algorithmic}
\end{algorithm}

\section{Simulation}\label{sec:results}
In each simulation run, the following procedure is adopted:

\begin{enumerate}[label=Step\arabic*:]
\item Random UEs placement and channel generation: The positions of $K$ UEs are independently and uniformly generated within a predefined coverage area. Based on these positions, the corresponding channels are generated. 
\item Codebook construction: A codebook of $K$ RIS phase shift configurations is constructed by assigning all possible $b$-bit quantized phase values to the $N$ PRUs. Each codeword corresponds to a unique $N$-dimensional phase vector.
\item SNR loss matrix calculation: For each pair of codewords in the codebook, the relative SNR degradation resulting from applying one configuration in place of another is calculated according to the defined loss metric.
\item Index assignment strategies: Different index assignment strategies are applied to the codebook.
\item Loss averaging: For each assignment strategy, the expected SNR loss due to single-bit feedback errors is estimated by averaging the relative SNR degradation over all codeword pairs whose indices differ by exactly one bit.
\end{enumerate}

Unless otherwise specified, all results are averaged over multiple ($>100$) independent random UE deployments and channel realizations to ensure statistical reliability. The default system parameters, unless varied for specific experiments, are summarized as Table \ref{tab:sim_params}.

\begin{table}[h]
\centering
\caption{Default simulation parameters}
\begin{tabular}{p{1cm}p{3.5cm}p{3cm}}
\hline
{Parameter}           & {Description}              & {Default Value} \\
\hline
$M$         & Number of BS antennas              & 16 \\
$N$         & Number of RIS PRUs                 & 256 \\
$K$         & Number of codewords (RIS configs) or UEs & 256 \\
$b$         & Quantization bits per PRU          & 8 \\
$P$         & Transmit power                     & 1 W \\
$\sigma^2$  & Noise power                        & $-90$ dBm \\
$q$         & Feedback bit error probability     & $10^{-3}$ \\
$L$         & Number of Monte Carlo runs         & 100 \\
UE area     & UE deployment region               & $50\,\mathrm{m} \times 50\,\mathrm{m}$ square \\
Channel & Channel path loss and fading   &  Rayleigh \\
\hline
\end{tabular}
\label{tab:sim_params}
\end{table}

The mapping from BSC SNR to bit error probability $q$ follows standard results for BPSK modulation over AWGN channels (see Appendix B for details). Table \ref{tab:bsc} lists the bit error rates $q$ corresponding to each BSC SNR value used in our simulations.

\begin{table}[h]
\centering
\caption{Bit error rate $q$ versus BSC SNR (dB)}
\begin{tabular}{c c}
\hline
{BSC SNR (dB)} & {Bit Error Rate $q$} \\
\hline
0  & 0.07865 \\
1  & 0.05628 \\
2  & 0.03751 \\
3  & 0.02288 \\
4  & 0.01250 \\
5  & 0.00595 \\
6  & 0.00239 \\
8  & $1.91 \times 10^{-4}$\\
12 & $9.01 \times 10^{-9}$ \\
\hline
\end{tabular}
\label{tab:bsc}
\end{table}

\subsection{Experiment I: SIMO-RIS SNR Loss Comparison — TSP vs. Natural and Random Indexing}

\subsubsection{Simulation Setup}
We consider a SIMO-RIS system with $N=256$ RIS PRUs and $K=256$ codewords, corresponding to serving 256 UEs,  each phase shift is quantized using $b=8$ bits (i.e., $\Delta\theta = 2\pi/256$). The BS is equipped with $M=16$ antennas. The feedback channel follows a BSC model, with SNR ranging from 0 dB to 12 dB in 4 dB steps, corresponding to varying single-bit error rates $q$.

\begin{figure}[htbp]
\centering
\includegraphics[width=0.5\textwidth]{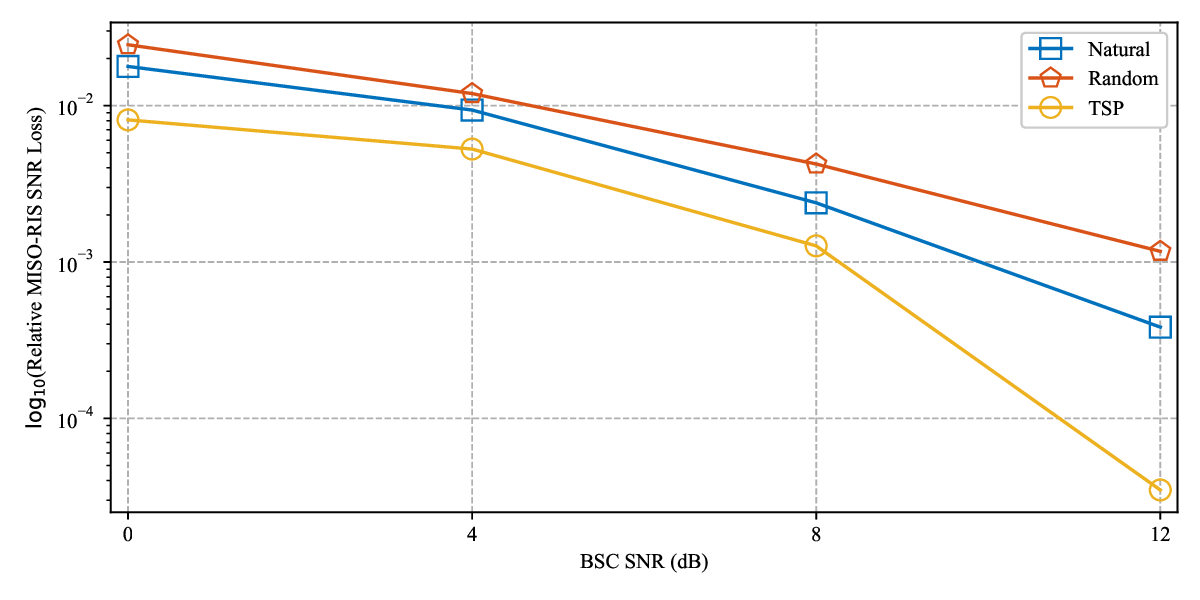}
\caption{Average relative SNR loss versus BSC SNR for three indexing strategies: natural, random, and TSP-based ($N=256$, $b=8$, $K=256$, $M=16$).}
\label{fig:snr_loss_new}
\end{figure}

\subsubsection{Results}

As shown in Fig. \ref{fig:snr_loss_new}, the TSP-based index assignment consistently achieves lower average SNR loss compared to natural and random mappings. The performance gap widens as the SNR decreases, demonstrating the robustness of the TSP strategy in minimizing performance degradation due to index mismatch. In contrast, natural and random assignments exhibit noticeably higher SNR loss, particularly in scenarios with severe feedback errors.

\subsection{Experiment II: Impact of System Parameters}

\subsubsection{Simulation Setup}

The BSC SNR is swept over $\{0, 1, 2, 3, 4, 5, 6\}$ dB. Unless otherwise specified, the default configuration uses $N=256$ RIS PRUs, quantization bits $b=8$, $K=256$ UEs and $M=16$ BS antennas. We vary the RIS size ($N\in\{64, 128, 256\}$), quantization level ($b=8, 10, 12$), and the number of BS antennas ($M=16, 32$). For each setting, we evaluate the average relative SNR loss under TSP-based index assignment.

\begin{figure}[htbp]
\centering
\includegraphics[width=0.5\textwidth]{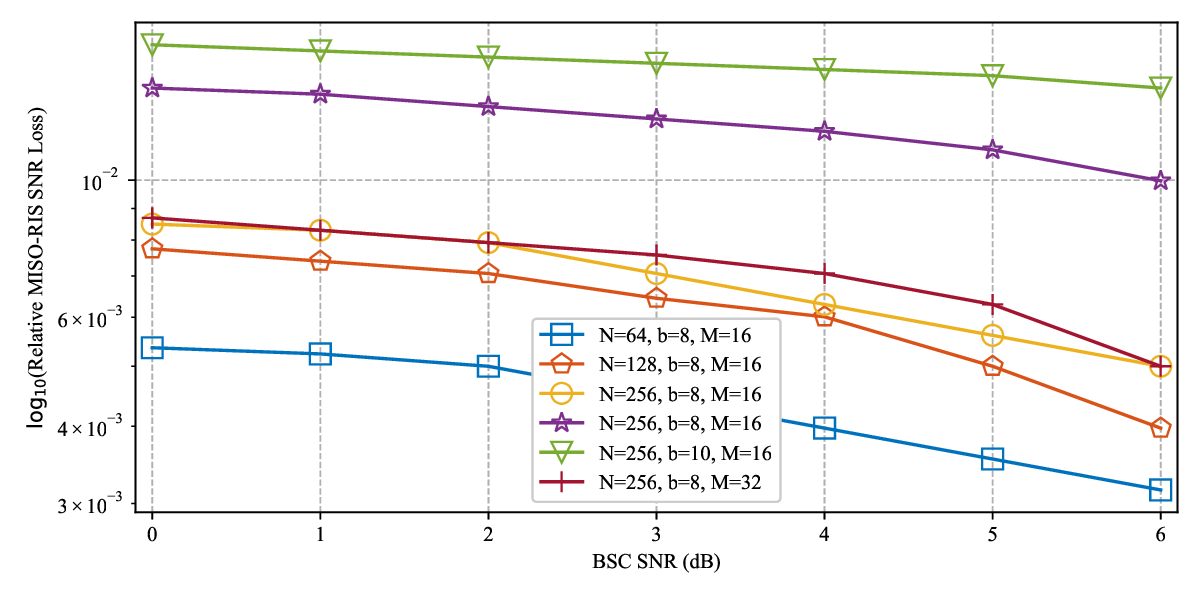}
\caption{Relative SNR loss under varying system parameters.}
\label{fig:param_sensitivity_new}
\end{figure}

\subsubsection{Results}

As shown in Fig.~\ref{fig:param_sensitivity_new}, the average SNR loss increases as either $N$ (the number of RIS PRUs) or $b$ (the phase quantization bits) grows, especially at low BSC SNR values where feedback errors are more pronounced. For instance, when $N=256$ and $b=12$, the SNR loss is significantly higher than for $N=64$ or lower $b$ values, illustrating the increased sensitivity to bit errors in larger and more finely quantized systems. Increasing the number of BS antennas $M$ from 16 to 32 (with other parameters fixed) leads to a slight rise in SNR loss, but its impact remains secondary compared to $N$ or $b$.

\subsection{Experiment  III: Ablation Study}
\subsubsection{Simulation setup}
Set $N=256$, $M=16$, $b=8$, $K=256$, BSC SNR$\approx 0\rm dB$. We enable/disable the following modules: (1) Layered candidate sets; (2) Adaptive $\mu$; (3) Dynamic $k_{\rm cate}$ reduction (adaptive $z$); (4) Edge-frequency filtering; (5) Final exhaustive refinement. 

\begin{figure}[htbp]
\centering
\includegraphics[width=0.5\textwidth]{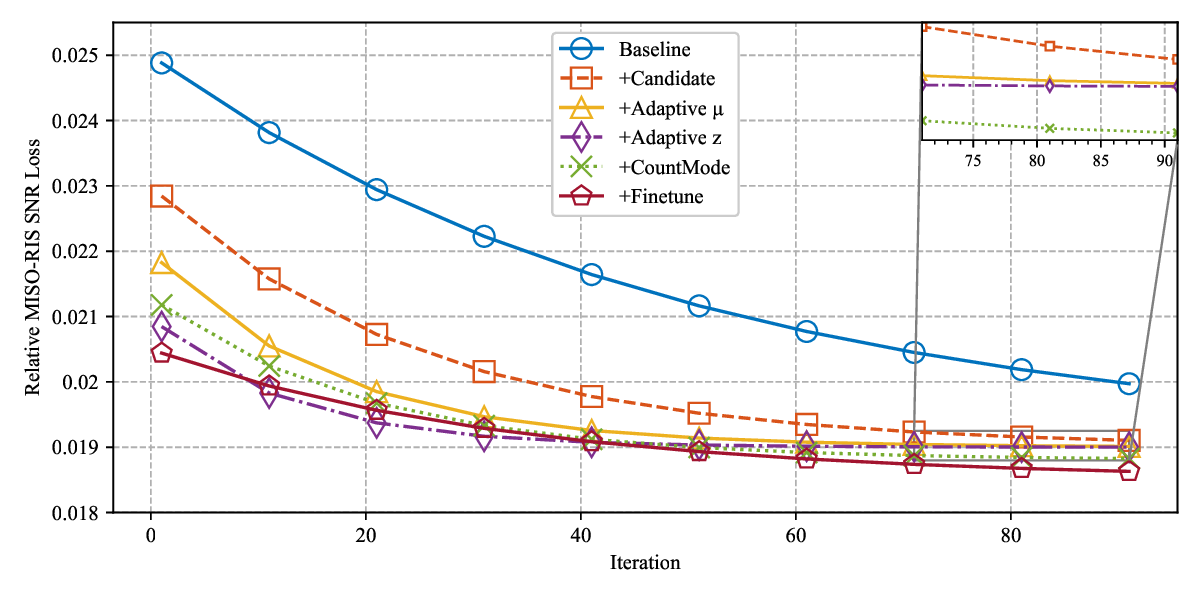}
\caption{Convergence curves over iterations for ablation experiments($N=256$, $M=16$, $b=8$, $K=256$, BSC SNR$\approx 0\rm dB$). }
\label{fig:ablation}
\end{figure}
\subsubsection{Results}
From Fig. \ref{fig:ablation}, we observe that layering accelerates convergence, adaptive $\mu$ boosts early progress, dynamic $k_{\rm cate}$(adaptive $z$) further speeds up, and edge-frequency filtering with final refinement yields the best final solution.

\subsection{Experiment III: Comparison with Other TSP Solvers}
\subsubsection{Simulation Setup}
Let $N=256$, $b=8$, $M=16$, and fix the number of codebooks at $K=256$ when investigating the impact of BSC SNR. The BSC SNR is swept over $\{0, 1, 2, 3, 4\}$ dB to evaluate solver robustness under varying feedback channel quality. In addition, to examine scalability with respect to the number of RIS PRUs, we sweep $N$ over $\{64, 128, 256, 512, 1024\}$ while fixing other parameters ($b=8$, $M=16$, $K=256$, BSC SNR$\approx 0\rm dB$).

We compare several TSP solvers, including LKH3~\cite{helsgaun2017extension}, Concorde~\cite{applegate2006concorde}, H-TSP~\cite{pan2023h}, GLOP~\cite{ye2024glop}, POMO~\cite{kwon2020pomo}, ELG~\cite{gao2023towards}, 2-opt, and the proposed method. For each method and each setting of BSC SNR or RIS PRUs, we record both the average relative MISO-RIS SNR loss and the solver execution time.

The open-source codes and license information used for benchmarking are listed in Appendix C.  
For computing platform specifications, refer to Appendix D.

\begin{figure}[htbp]
\centering
\includegraphics[width=0.5\textwidth]{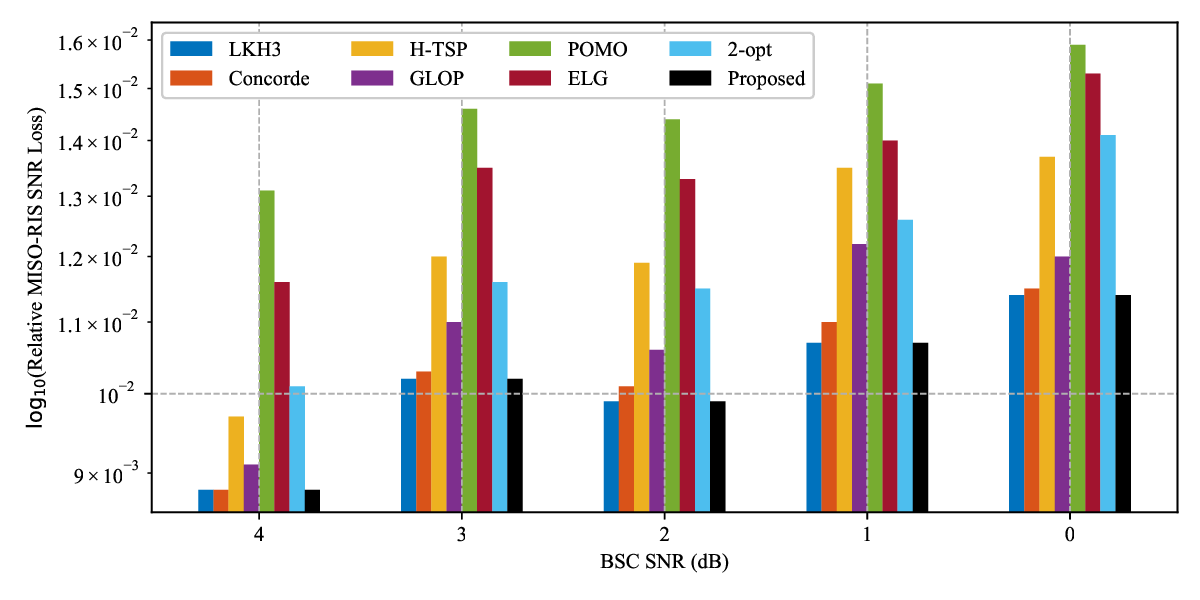}
\caption{Average SNR loss of different solvers as a function of BSC SNR ($b=8$, $M=16$, $K=256$, $N=256$).}
\label{fig:solver_perf_snr}
\end{figure}

\begin{figure}[htbp]
\centering
\includegraphics[width=0.5\textwidth]{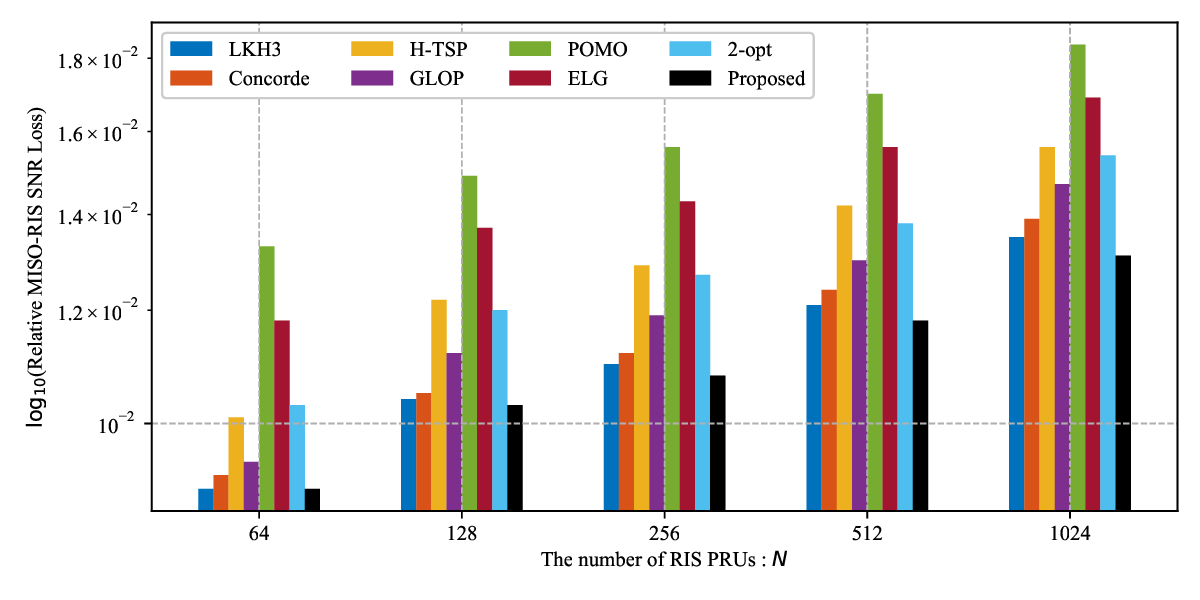}
\caption{Average SNR loss of different solvers as a function of the number of RIS PRUs ($b=8$, $M=16$, $K=256$, BSC SNR$\approx 0\rm dB$).}
\label{fig:solver_perf_n}
\end{figure}

\begin{figure}[htbp]
\centering
\includegraphics[width=0.5\textwidth]{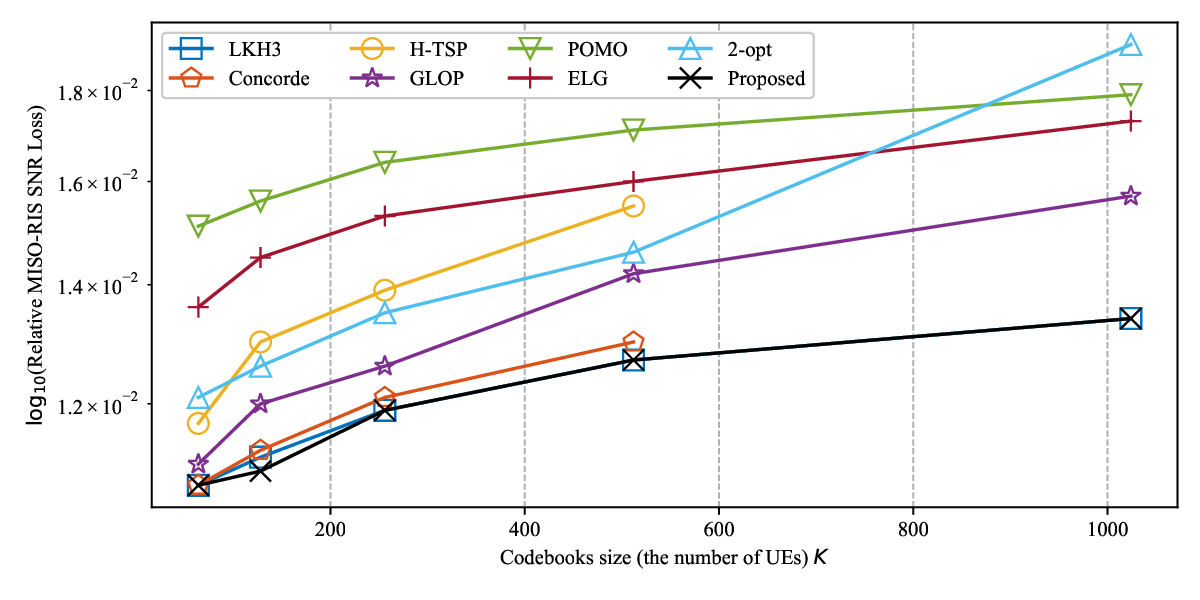}
\caption{Comparison of relative SNR loss for different solvers as a function of the number of codebooks $K$ (up to 1024), ($b=8$, $M=16$, $N=256$, BSC SNR$\approx 0\rm dB$).}
\label{fig:solver_perf_largeK}
\end{figure}

\begin{figure}[htbp]
\centering
\includegraphics[width=0.5\textwidth]{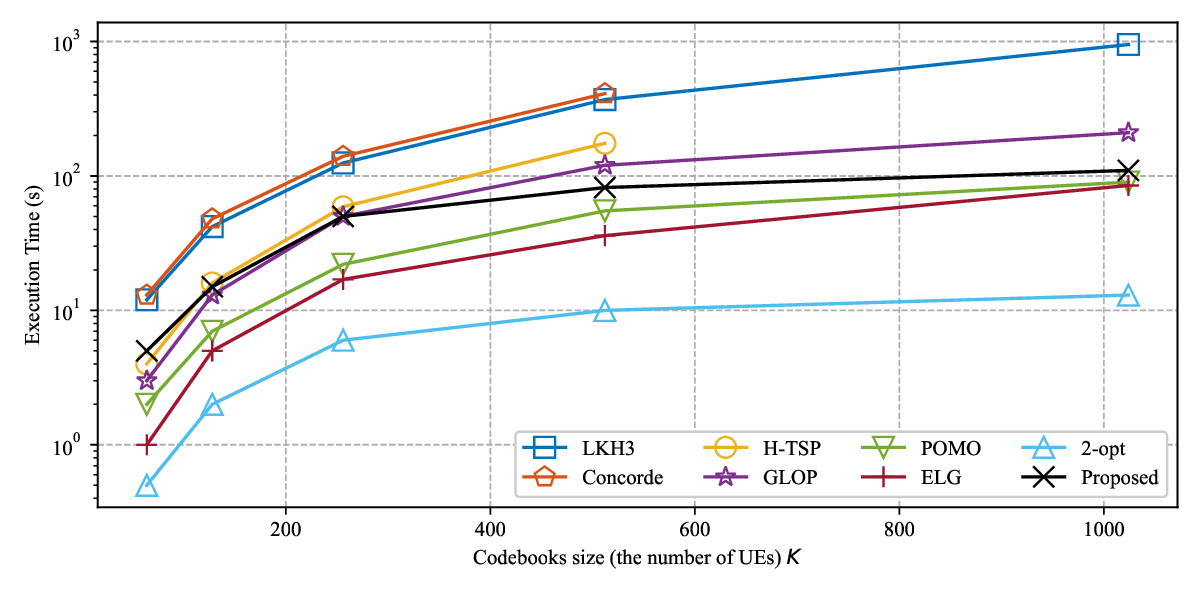}
\caption{Execution time of different solvers as a function of the number of codebooks $K$ (up to 1024), ($b=8$, $M=16$, $N=256$, BSC SNR$\approx 0\rm dB$).}
\label{fig:solver_time_largeK}
\end{figure}

\begin{figure*}[htbp]
\centering
\includegraphics[width=1.0\textwidth]{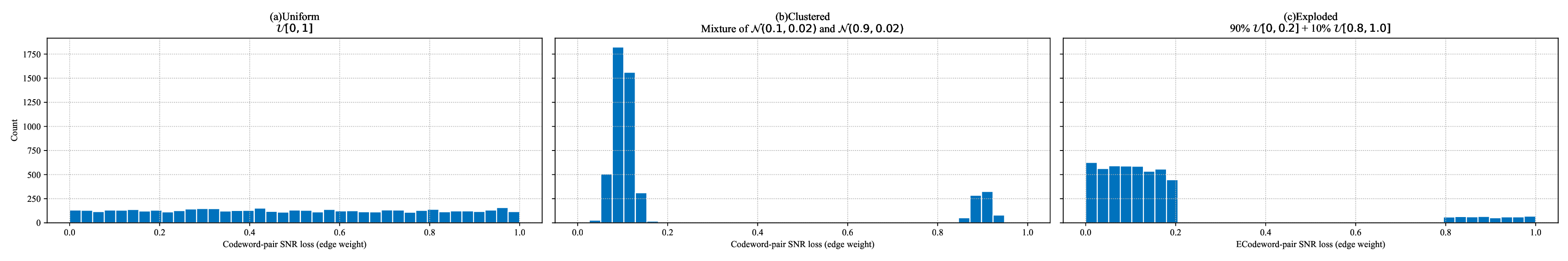}
\caption{Visualization of codeword-pair (edge weight) distributions: clustered and exploded.}
\label{fig:pair_dist}
\end{figure*}

\begin{table*}[htbp]
\centering
\caption{Performance comparison of TSP solvers for RIS codebook mapping (SNR loss and execution time, uniform distribution)}
\label{tab:tsp_uniform_ris}
\resizebox{1.03\textwidth}{!}{
\begin{tabular}{lcccccccccc}
\hline
\multirow{2}{*}{Method} & \multicolumn{2}{c}{Uniform $K=64$} & \multicolumn{2}{c}{Uniform $K=128$} & \multicolumn{2}{c}{Uniform $K=256$} & \multicolumn{2}{c}{Uniform $K=512$} & \multicolumn{2}{c}{Uniform $K=1024$} \\
\cmidrule(lr){2-3} \cmidrule(lr){4-5} \cmidrule(lr){6-7} \cmidrule(lr){8-9} \cmidrule(lr){10-11}
 & SNR Loss & Time (s) & SNR Loss & Time (s) & SNR Loss & Time (s) & SNR Loss & Time (s) & SNR Loss & Time (s) \\
\hline
LKH3\cite{helsgaun2017extension}      & \textbf{0.0108} & 12   & 0.0112 & 42   &  \textbf{0.0119} & 125   &  \textbf{0.0127} & 370   &  \textbf{0.0134} & 950   \\
Concorde\cite{applegate2006concorde}  &  \textbf{0.0108} & 13   & 0.0113 & 48   & 0.0121 & 140   & 0.0130 & 410   & - & -  \\
H-TSP\cite{pan2023h}                  & 0.0117 & 4    & 0.0130 & 16   & 0.0139 & 59    & 0.0155 & 175   & - & -   \\
GLOP\cite{ye2024glop}                 & 0.0111 & 3    & 0.0120 & 13   & 0.0126 & 50    & 0.0142 & 120   & 0.0157 & 210   \\
POMO\cite{kwon2020pomo}               & 0.0151 & 2    & 0.0156 & 7    & 0.0171 & 22    & 0.0164 & 55    & 0.0173 & 90    \\
ELG\cite{gao2023towards}              & 0.0136 & 1    & 0.0145 & 5    & 0.0153 & 17    & 0.0160 & 36    & 0.016 & 85    \\
2-opt                                 & 0.0121 & 0.5  & 0.0126 & 2    & 0.0135 & 6     & 0.0146 & 10    & 0.0191 & 13    \\
3-opt                                 & 0.014 & 1    & 0.014 & 4    & 0.015 & 12    & 0.016 & 30    & 0.017 & 65    \\
Greedy                                & 0.016 &  \textbf{0.1}  & 0.016 &  \textbf{0.2}  & 0.018 &  \textbf{0.8}   & 0.019 &  \textbf{2}     & 0.0246 &  \textbf{4}    \\
GA                                    & 0.014 & 1.5  & 0.015 & 6    & 0.016 & 22    & 0.017 & 65    & 0.017 & 140   \\
ACO                                   & 0.014 & 2.3  & 0.015 & 9    & 0.016 & 35    & 0.016 & 95    & 0.017 & 200   \\
Proposed                              &  \textbf{0.0108} & 5    &  \textbf{0.0110} & 15   &  \textbf{0.0119} & 50    &  \textbf{0.0127} & 82    &  \textbf{0.0134} & 110   \\
\hline
\end{tabular}
}
\end{table*}

\begin{table*}[htbp]

\caption{Cross-distribution generalization: TSP solvers on clustered/exploded codeword-pair distributions}
\label{tab:tsp_cross_ris}
\begin{tabular}{lcccccccccccc}
\toprule
\multirow{2}{*}{Method} & \multicolumn{2}{c}{Clustered $K=64$} & \multicolumn{2}{c}{Exploded $K=64$} & \multicolumn{2}{c}{Clustered $K=128$} & \multicolumn{2}{c}{Exploded $K=128$} & \multicolumn{2}{c}{Clustered $K=256$} & \multicolumn{2}{c}{Exploded $K=256$} \\
\cmidrule(lr){2-3} \cmidrule(lr){4-5} \cmidrule(lr){6-7} \cmidrule(lr){8-9} \cmidrule(lr){10-11} \cmidrule(lr){12-13}
 & SNR Loss & Time & SNR Loss & Time & SNR Loss & Time & SNR Loss & Time & SNR Loss & Time & SNR Loss & Time \\
\midrule
LKH3\cite{helsgaun2017extension}     & \textbf{0.01133} & 6.4s  & \textbf{0.01067} & 6.1s  & 0.01210 & 33.2s & 0.01188 & 32.7s & 0.01342 & 2.8m & 0.01320 & 2.8m \\
Concorde \cite{applegate2006concorde}     & 0.01144 & 9.5s  & 0.01111 & 9.1s  & 0.01232 & 39.0s & 0.01221 & 38.5s & †OOM     & -    & †OOM     & -    \\
H-TSP\cite{pan2023h}   & 0.01287 & 3.1s  & 0.01199 & 3.2s  & 0.01463 & 18.1s & 0.01386 & 17.2s & †OOM     & -    & †OOM     & -    \\
GLOP\cite{ye2024glop}         & 0.01232 & 2.9s  & 0.01177 & 2.8s  & 0.01408 & 11.2s & 0.01342 & 10.8s & 0.01485 & 41.1s & 0.01441 & 39.0s \\
POMO\cite{kwon2020pomo}        & 0.01661 & 0.61s & 0.01584 & 0.61s & 0.01749 & 2.3s  & 0.01672 & 2.3s  & 0.01837 & 10.8s & 0.01760 & 10.8s \\
ELG\cite{gao2023towards}    & 0.01595 & 0.17s & 0.01507 & 0.17s & 0.01672 & 0.61s & 0.01606 & 0.61s & 0.01760 & 2.9s  & 0.01694 & 2.9s  \\
2-opt       & 0.01364 & 0.09s & 0.01287 & 0.09s & 0.01496 & 0.41s & 0.01441 & 0.39s & 0.01617 & 2.1s  & 0.01562 & 2.1s  \\
3-opt       & 0.01298 & 0.39s & 0.01243 & 0.41s & 0.01419 & 1.3s  & 0.01397 & 1.2s  & 0.01507 & 9.1s  & 0.01474 & 8.9s  \\
Greedy      & 0.01551 & \textbf{0.027s}& 0.01452 & \textbf{0.029s}& 0.01617 & \textbf{0.080s}& 0.01773 & \textbf{0.083s}& 0.01916 & \textbf{0.22s} & 0.02672 & \textbf{0.22s} \\
GA          & 0.01606 & 2.41s & 0.01518 & 2.40s & 0.01705 & 8.1s  & 0.01639 & 8.1s  & 0.01782 & 24.2s & 0.01727 & 24.5s \\
ACO         & 0.01265 & 3.6s  & 0.01199 & 3.6s  & 0.01364 & 13.7s & 0.01342 & 13.7s & 0.01507 & 56.0s & 0.01485 & 56.0s \\
Proposed    & \textbf{0.01133} & 3.72s & 0.01089 & 3.72s & \textbf{0.01177} & 14.2s & \textbf{0.01166} & 14.2s & \textbf{0.01298} & 68.7s & \textbf{0.01287} & 68.7s \\
\bottomrule
\end{tabular}
\vspace{1mm}
\noindent\footnotesize{†OOM: out-of-memory}
\end{table*}

\subsubsection{Results}
As shown in Fig.~\ref{fig:solver_perf_snr}, the proposed method consistently achieves the lowest relative SNR loss across all tested BSC SNR values, outperforming both classical and learning-based TSP solvers. The performance gap becomes more pronounced as the BSC SNR decreases (i.e., feedback errors become more likely), highlighting the robustness of the proposed assignment strategy under harsh channel conditions. Notably, some learning-based methods (e.g., POMO, ELG) and fast heuristics (e.g., 2-opt) suffer significantly higher SNR loss in the low-SNR regime.

Fig. \ref{fig:solver_perf_n} further demonstrates the scalability of all methods with respect to the number of RIS PRUs. As $N$ increases, all solvers exhibit increased SNR loss, which is expected due to the higher problem complexity. However, the proposed method continues to deliver the best SNR performance.

In addition, as shown in Fig. \ref{fig:solver_perf_largeK} and \ref{fig:solver_time_largeK}, the superiority of the proposed method becomes even more pronounced in large-scale scenarios ($K$ up to 1024).  Simpler heuristics such as 2-opt have the lowest execution time, but their SNR loss is significantly worse. While most baseline solvers suffer from either rapidly increasing SNR loss (e.g., 2-opt, H-TSP, POMO) or excessive computation time (e.g., LKH3, Concorde), the proposed approach maintains the lowest SNR loss across all tested values of $K$ and scales reasonably in terms of runtime. This demonstrates that the proposed method is highly suitable for practical systems with massive codebook sizes, where both solution quality and computational efficiency are critical. 

H-TSP and Concorde fail for large $K=1024$ due to excessive computational and memory demands.

\subsection{Experiment IV: Cross-Distribution Generalization of TSP Solvers}

\subsubsection{Simulation Setup}

In this experiment, we focus on evaluating the cross-distribution generalization ability of various TSP solvers for the RIS codebook index assignment problem. In practical RIS systems, the codeword-pair (edge) weights may exhibit different statistical structures, which can impact solver robustness and SNR performance. We specifically consider two challenging non-uniform distributions:

\begin{itemize}
\item {Clustered:} Most edge weights are drawn from a low-variance Gaussian $\mathcal{N}(0.1, 0.02)$, while a minority are sampled from a higher-mean Gaussian $\mathcal{N}(0.9, 0.02)$, representing a scenario where most codeword pairs are “similar” but a few are very different.
\item {Exploded:} 90\% of weights are small ($\mathcal{U}[0, 0.2]$), while 10\% are large ($\mathcal{U}[0.8, 1.0]$), creating a highly bimodal or “exploded” structure, as often arises in quantized phase systems with sparse error peaks.
\end{itemize}

For each scenario, TSP instances are generated for $K=64, 128, 256$, with results averaged over 30 random seeds for statistical reliability. The codeword-pair weight distributions are visualized in Fig.\ref{fig:pair_dist}. We benchmark a comprehensive set of state-of-the-art and classical TSP solvers, including LKH3, Concorde, H-TSP, GLOP, POMO, ELG, 2-opt, 3-opt, Greedy, Genetic Algorithm (GA), Ant Colony Optimization (ACO), and the proposed solver. For each solver and setting, both the achieved objective (relative MISO-RIS SNR loss) and execution time are recorded (see Table \ref{tab:tsp_uniform_ris} and \ref{tab:tsp_cross_ris}).

\subsubsection{Results}
As observed in Table~\ref{tab:tsp_uniform_ris}, advanced solvers including LKH3~\cite{helsgaun2017extension}, Concorde~\cite{applegate2006concorde}, H-TSP~\cite{pan2023h}, GLOP~\cite{ye2024glop}, and the proposed method achieve the lowest relative SNR loss for all codebook sizes, with their performance remaining tightly clustered (SNR loss typically within $0.001$ to $0.002$ of each other) even as $K$ increases from 64 to 1024. Notably, the proposed method consistently matches or slightly outperforms LKH3 and H-TSP in SNR loss, while being significantly more efficient as the problem scale increases. For example, at $K=1024$, the proposed method completes in 110 seconds, which is nearly an order of magnitude faster than LKH3 (950 s). 

Heuristic and metaheuristic algorithms—including 2-opt, 3-opt, Greedy, GA, ACO, as well as learning-based solvers such as POMO~\cite{kwon2020pomo} and ELG~\cite{gao2023towards}—exhibit much shorter runtimes, generally completing within a few seconds or less for $K \leq 256$. However, this speed comes at the cost of a persistent and increasing SNR loss gap: while for small $K$ ($K=64$) their SNR loss is only about $0.002$–$0.003$ worse than optimal, for $K=1024$ the gap widens to $0.004$–$0.011$, with worst-case losses exceeding $0.0246$ (Greedy). This trend indicates that, as the codebook size grows, the simple or local search methods struggle to match the solution quality of advanced or tailored solvers, highlighting the importance of robust optimization techniques in large-scale applications.

Table~\ref{tab:tsp_cross_ris} reports the cross-distribution generalization results for all solvers under clustered and exploded (bimodal) codeword-pair weight scenarios, which emulate practical non-uniformities in the RIS codebook structure due to hardware constraints or environment-induced quantization artifacts. Here, all algorithms experience an increase in SNR loss relative to the uniform case, reflecting the greater challenge posed by highly variable edge penalties. Nevertheless, the relative ranking among solvers is largely preserved: the proposed method and LKH3 remain the most robust across both distributions and all tested $K$. For example, in the most challenging exploded case with $K=256$, the proposed algorithm yields an SNR loss of 0.01298 and 0.01287, outperforming all other baselines and remaining close to LKH3 (0.01342 and 0.01320), which is the next best. For $K=128$ and $K=64$, similar trends are observed, with both advanced solvers suffering only minor performance degradation as the edge weight distribution becomes more irregular.

Metaheuristic and learning-based solvers, such as POMO and ELG, exhibit pronounced SNR loss inflation in non-uniform settings, with worst-case losses up to $0.01760$ (POMO, exploded $K=256$), a gap of more than $0.004$ relative to the top-performing methods. Fast heuristics like Greedy, 2-opt, and 3-opt are even more sensitive to distribution irregularity, sometimes losing $20\%$ or more SNR compared to the optimal. Genetic and evolutionary approaches (GA, ACO) display intermediate behavior, performing moderately well for clustered distributions but still failing to close the gap under “exploded” penalty regimes. On the computational side, the proposed method demonstrates stable, polynomially-scaling runtime across all settings, completing all large-scale tasks without out-of-memory or time limit issues. In contrast, exact solvers such as Concorde and H-TSP become impractical for $K=256$ (and impossible for $K=512,1024$), as indicated by the OOM entries in Table~\ref{tab:tsp_cross_ris}.

\section{Conclusion and Future Work}\label{sec:conclusion}
This paper proposed a robust index assignment framework for RIS codebook configuration under feedback errors, formulating the problem as a TSP to minimize SNR loss from likely index mismatches. A scalable three-phase solver was developed and shown to achieve near-optimal robustness with low complexity, outperforming baseline and state-of-the-art methods in various scenarios.

Future work will address extensions to multi-bit error regimes, adaptive and online mapping under dynamic channels, and practical system integration.

\bibliographystyle{IEEEtran}
\bibliography{references.bib}

\appendix

\section*{A: Proof: The proposed TSP solver has polynomial complexity.}
\label{sec:complexity}
Let $K$ denote the number of codewords (i.e., nodes or cities in the TSP formulation), $T$ the number of outer iterations in the Fuzzy Concatenation Phase (with worst-case $T=O(K)$), and $|\Theta|$ the size of the candidate set for the next codeword (typically $O(K)$).

\medskip
\noindent
\textbf{Claim:} The overall worst-case time complexity of the proposed algorithm is $O(K^4)$.

\medskip
\noindent
\textbf{Proof:}
\begin{enumerate}[label=\textbf{ }]

\item \textit{$\lozenge$ Provision Phase:} \\
In this initialization phase, each codeword is processed exactly once to assign and sort its neighbor layers according to pairwise perturbation values. For each of the $K$ codewords, creating these layers involves sorting and assignment over the remaining $K-1$ codewords, resulting in $O(K)$ operations per codeword. Therefore, the total complexity for this phase is $O(K^2)$. As this is a one-time preprocessing step, its impact is negligible compared to the subsequent phases when $K$ is large.

\item \textit{$\lozenge$ Shotgun Phase:} \\
This phase is the main candidate generation stage. The outer loop is executed $3K^2$ times, each representing a sampled Hamiltonian path through the codewords. For each sampled path:
\begin{itemize}
    \item The path is constructed in $K$ steps (one codeword added per step).
    \item At each step, the candidate set $\Theta$ is traversed to select the next codeword, and $|\Theta|=O(K)$ in the worst case, as any unvisited codeword can be a candidate.
\end{itemize}
Thus, the total sampling complexity is $O(3K^2 \times K \times K) = O(K^4)$. 
After all samples are generated, the algorithm sorts the $3K^2$ sampled paths to retain the best $k_{\text{shot}}$ (as described in the main text). Sorting $O(K^2)$ nodes requires $O(K^2 \log K)$ operations, which is dominated by the $O(K^4)$ complexity from the path generation step for large $K$.

\item \textit{$\lozenge$ Fuzzy Concatenation Phase:} \\
This phase iteratively refines the solution pool using historical edge-frequency statistics. For each of the $T$ outer iterations (with $T=O(K)$ in the worst case):
\begin{itemize}
    \item $n_{\rm cate} = 200K$ new candidate routes are generated.
    \item Each route construction involves $K$ steps, and at each step, up to $O(K)$ candidates are examined based on pair statistics and greedy/weighted selection.
\end{itemize}
Thus, the complexity per iteration is $O(200K \times K \times K) = O(K^3)$. Over $T=O(K)$ iterations, the total cost is $O(T \times K^3) = O(K^4)$ in the worst case. \\
In addition, after each iteration, $n_{\rm cate}$ routes are sorted to select the best $k_{\rm cate}$, costing $O(K \log K)$ per iteration, which remains negligible compared to the path construction step.

Auxiliary operations such as removal of repeated codewords, fine-tuning of the last few codewords in a path (small exhaustive search), and pair-count updates contribute at most $O(K^2)$ or $O(K^3)$ per iteration, and do not increase the overall asymptotic complexity.

The algorithm includes a dynamic mode-switching mechanism to avoid local optima, where the search mode (Normal, Intermediate, or Upper Limit) is changed only if no improvement is observed for several iterations. The number of such switches is bounded by a small constant (independent of $K$) and does not affect the overall complexity. In practice, the number of Fuzzy Concatenation iterations $T$ rarely reaches the worst-case $O(K)$ upper bound.
\end{enumerate}

\medskip
\noindent{Summary:} The computational complexity of the proposed three-phase heuristic algorithm can be characterized as follows: the Provision Phase incurs complexity $O(K^2)$, which is negligible for large codebook sizes $K$; the Shotgun Phase and Fuzzy Concatenation Phase each have complexity $O(K^4)$. Summing these individual complexities, the algorithm's total worst-case computational complexity is $O(K^4)$. Since this complexity is polynomial in $K$, the algorithm is computationally feasible and practical for moderate to large-scale RIS systems. \hfill$\blacksquare$

\section*{B: Bit Error Rate Computation in Simulation}
\label{sec:biterror}
Given an BSC SNR in dB, we first convert it to linear scale as
\begin{equation}
\mathrm{SNR}_{\rm lin} = 10^{\mathrm{SNR}_{\rm dB}/10}.
\end{equation}
For BPSK modulation over an AWGN channel, the bit error rate (BER) is given by
\begin{equation}
q = Q(\sqrt{2\mathrm{SNR}_{\rm lin}}) = \frac{1}{2}\mathrm{erfc}(\sqrt{\mathrm{SNR}_{\rm lin}}),
\end{equation}
where $Q(x)=\tfrac{1}{2}\mathrm{erfc}(x/\sqrt{2})$ is the tail probability of the standard normal distribution.

In practice, for small $q$ (i.e., in the high BSC SNR regime), the probability of multiple simultaneous bit errors is negligible, so only single-bit flips dominate the average loss calculation. Hamming-distance-1 neighbors correspond exactly to adjacent nodes on the TSP path. Through this mapping, any 1-bit error—the most probable error type—results in minimal RIS configuration perturbation, thereby improving link robustness. However, if the feedback error rate is high, multi-bit errors become non-negligible, or the index distribution is non-uniform, a more general probability-weighted QAP formulation should be adopted instead of the TSP simplification. This modeling assumption has been widely applied in codebook index allocation and distributed signal processing.

\section*{C: List of Open-Source Codes for Benchmarks}
\label{sec:benchmarks}
\begin{table}[h]
\centering
\caption{List of links and licenses for the codes used}
\begin{tabular}{p{1.5cm}|p{4cm}|p{2cm}}
\hline 
Resource & Link & License \\
\hline 
LKH3~\cite{helsgaun2017extension} & \url{http://webhotel4.ruc.dk/-keld/research/LKH-3/} & Research-usable \\
Concorde~\cite{applegate2006concorde} & \url{https://github.com/jvkersch/pyconcorde} & BSD 3-Clause License \\ 
H-TSP~\cite{pan2023h} & \url{https://github.com/Learning4Optimization-HUST/H-TSP} & Research-usable \\
GLOP~\cite{ye2024glop} & \url{https://github.com/henry-yeh/GLOP} & MIT License \\
POMO~\cite{kwon2020pomo} & \url{https://github.com/yd-kwon/POMO/tree/master/NEW_py_ver} & MIT License \\
ELG~\cite{gao2023towards} & \url{https://github.com/gaocrr/ELG} & MIT License \\
\hline
\end{tabular}
\end{table}

\section*{D: Simulation Platform}
\label{sec:platform}
All experiments were conducted on a high-performance computing cluster with the following specifications:
\begin{itemize}
\item {CPU:} 2 × Intel Xeon Scalable Cascade Lake 8168 (2.7GHz, 24 cores)
\item {Memory:} 1.5 TB DDR4 ECC REG 2666
\item {GPU:} 16 × NVIDIA Tesla V100 (8 GPUs per node, total 16 GPUs)
\end{itemize}
The GPU resources were utilized primarily for accelerating learning-based methods, including GLOP~\cite{ye2024glop}, POMO~\cite{kwon2020pomo}, and ELG~\cite{gao2023towards}. All these frameworks fall under the category of neural combinatorial optimization (NCO), leveraging deep learning and neural network models to efficiently handle large-scale combinatorial optimization problems such as TSP.

\vfill

\end{document}